\documentstyle{article}
\begin{document}
{\rm
\ \\
\rightline{hep-ph/9712288}
}
\begin{center}
{\Large \bf Double-Flavor Oscillations}\\
\ \\
{\Large \bf and Properties of Heavy Mesons}
\footnote{Presented at the {\it 12th International Workshop
 on High Energy Physics and Quantum Field Theory}, Samara,
 Russia, \mbox{4--10 September 1997}. To be published in the
 Proceedings.}

\vspace{4mm}

Ya.I.Azimov\\
Petersburg Nuclear Physics Institute\\
Gatchina, St.Petersburg, 188350, Russia\\
\end{center}

\begin{abstract}
Phenomenon of coherent double-flavor oscillations is discussed. They
can arise when heavy flavored neutral mesons decay with production
of neutral kaons. The oscillations may be used to study detailed
properties of the heavy mesons and their decays. They may give
new insight into the problem of $CP$-violation. In particular,
they provide the only known way to the complete unambiguous
measurements of $CP$-violating parameters for neutral $B$-meson
decays. The corresponding experiments seem to be hard, but
possible at hadron facilities.

\end{abstract}

Oscillations in decays of flavored neutral mesons is now a well-known
phenomenon. They are induced by the meson-antimeson mixing due to
flavor non-conservation. The strangeness oscillations in propagation
of neutral kaons provide the best studied example. Similar
oscillations of beauty have been observed for $B_d,\overline B_d$
mesons ($B_d$ is $\bar bd$). Flavor oscillations should exist also
for $B_s, \overline B_s$ and $D^0, \overline D^0$ ($B_s$ and $D^0$
are $\bar bs$ and $c\bar u$ correspondingly), and common belief is
that they do exist, though not measured yet.

Some years ago I suggested a new possible phenomenon in decays of
heavy neutral mesons, many-flavor oscillations~\cite{aJ,az}. Indeed,
such a meson oscillates itself, then it can decay, e.g., with
producing neutral kaons which demonstrate oscillations of
strangeness. Non-trivial is that the initial flavor oscillations
and the final strangeness oscillations may be coherent.

A simple example is given~\cite{aJ,az} by the primary decay
\begin{equation}
B_d(\overline B_d)\to J/\psi K^0(\overline K^0)\,, \end{equation}
followed by secondary decays of the kaon
\begin{equation}
K^0(\overline K^0)\to\pi\pi\;\;\;\;{\rm or}\;\;\;\;
K^0(\overline K^0)\to\pi^{\mp} l^{\pm}\nu_l(\bar\nu_l)\,.
\end{equation}
At the quark level this cascade is induced by the quark processes
$\bar b\to\bar cW^+,\, W^+\to c\bar s,\,\bar s\to\bar uW^+$ and
charge-conjugate.  We see that only transitions
\begin{equation} B_d\to K^0\,,\;\;\; \overline B_d\to\overline
K^0 \end{equation}
are possible in decay (1), while transitions $B_d\to\overline
K^0,\,\overline B_d\to K^0$ cannot go \footnote{The situation is
just opposite if the same cascade is initiated by $B_s$, see
ref.\cite{ad}.}. Therefore, $K$-meson content of the state generated
by the $B$-meson decay is unambiguously determined by the initial
$B$-meson state and its evolution up to the decay. The later
$K$-evolution appears to be coherent continuation of $B$-evolution
before the decay.

Such double coherence has many analogues in other physical phenomena.
Analogy with propagation of the polarized light through optically
active substances (rotating its polarization plane) was discussed
in ref.~\cite{az}. Analogy with the famous quantum effect of
Einstein-Podolsky-Rosen~\cite{epr} was emphasized in ref.~\cite{ks}.

Coherence of the secondary kaon evolution with processes at the
previous stage(s) of the cascade decay allows to use the well-known
kaon properties for studying unknown properties of heavy mesons
($B$-mesons in the above cascade) and their decays. To illustrate
how this approach can work we consider here some examples. We will
mainly trace the corresponding ideas and will not go into every
detail of calculations which can be found in the referred papers.
After that we will briefly discuss possibilities of the
corresponding experiments.

1. Let us first assume, for simplicity, exact $CP$-conservation and
compare a direct semileptonic decay of the neutral flavored meson
with similar decay of the secondary neutral kaon after decay (1).

Take, for definiteness, the pure $B$-meson initial  state. It is the
superposition of two eigenstates, $B_\pm$, with definite $CP
$-parities $\pm1$. They have definite masses and widths. Their
simple laws of evolution provide the relative factor $\exp(i\Delta
m_B t_B)$ after the lifetime $t_B$. Semileptonic decay amplitudes
for the eigenstates are the same (or have the opposite sign).
Therefore, their interference in any direct semileptonic decay gives
a contribution to the decay rate proportional to $\cos(\Delta m_B
t_B)$. Its study measures the absolute value  of the mass difference
$\Delta m_B$, but does not allow to find its sign, i.e. to determine
which of the states, $B_+$ or $B_-$, is heavier or lighter.

The situation is the same for all neutral flavored mesons. However,
special experiments for neutral kaons were capable to show that the
lighter kaon eigenstate is the (nearly) $CP$-even $K_S$, rapidly
decaying to two pions. This goal was achieved by observing kaon
semileptonic decays after coherent regeneration in foils (compilation
of results and references see in ref.\cite{pdt-old}). The
measurements appeared in agreement with later calculations in the
Standard Model. Such experiments are, surely, impossible for any
other neutral flavored mesons because of much shorter mean lifetimes.

Let us consider now the same semileptonic decay, but for the
secondary kaon, produced in the primary decay (1). Again, we start
with the pure $B$-meson and, for time being, assume $CP$ to conserve.
Then, considered in terms of eigenstates, decays (1), as explained in
ref.\cite{az}, have two branches
\begin{equation} B_+\to
J/\psi K_L\,,\;\;\;B_-\to J/\psi K_S\,, \end{equation}
each with a simple law of evolution in both lifetimes $t_B$ and
$t_K$ before and after decays (4). Contributions of the branches
to the secondary decay amplitude have the relative phase determined
by multiplication of two relative factors, $\exp(i\Delta m_B t_B)$
and $\exp(-i\Delta m_K t_K)$, coming from $B$- and $K$-evolutions.
The decay rate, after such double-flavor evolution, contains a term
proportional to $\cos( \Delta m_B t_B -\Delta m_K t_K)$. Thus,
contrary to the familiar single-flavor evolution, the cascade decay
with double-flavor evolution allows one to measure both the value and
sign of $\Delta m_B$, in respect to the known sign of $\Delta m_K$.

It is just the sign of $\Delta m_d$ that stays unknown today for
$B_d$, while its absolute value has been measured with good
precision in direct decays~\cite{pdt}. Cascade decays for $B_s
$-meson may appear even more useful~\cite{ad} since the large value
of $|\Delta m_s|$ generates too rapid oscillations in direct decays
which are hard to resolve.

2. $CP$-violation makes two more eigenstate transitions
\begin{equation}
B_+\to J/\psi K_S\,,\;\;\;B_-\to J/\psi K_L  \end{equation}
be possible as well~\cite{az}, in addition to transitions (4).
The cascade decay amplitude contains contributions from all four
transitions. Note, however, that only the same two transitions (3)
are admissible in terms of flavored states, independently of
conservation or violation of $CP$.

Essential problem under $CP$-violation is how to identify
eigenstates. $CP$-parities cease to be good quantum numbers, so one
needs to use some other properties. Neutral kaons are experimentally
identified by their mean lifetimes as short- or long-lived, $K_S$ or
$K_L$. This way cannot be useful for eigenstates of heavier mesons
since differences of their lifetimes are too small. Many authors
identify the eigenstates by their masses as heavy and light (e.g.,
$B_H,\,B_L$), but do not suggest how to relate such identification
with any other properties. The usual statement that $B_H$ is $CP
$-odd in the Standard Model (just as for kaons; see, e.g.,
ref.\cite{dan}) may appear uncertain because of uncertainty of the
$CP$-parity itself (it can be mode dependent, see discussion below).

Cascade decay (1) suggests another way of identification for the
eigenstates. According to ref.\cite{az}, we can define the states
$B_\pm$ as having approximate $CP$-parities $\pm1$, in the sense
that transitions (4) are more intense than transitions (5)
(considered then, by definition, as $CP$-suppressed). Similar
procedure is possible also for direct $B$-meson decays into final
states with definite $CP $-parities, but only if the difference of
eigenwidths is measurable (more details see in ref.\cite{az} for
cascade decays and in ref.\cite{ars} for direct decays). Such
identification by the approximate $CP$-parity reminds the earlier
identification of neutral kaon eigenstates as $\theta$- or
$\tau$-particles according to their decays into final states of two
or three pions, with opposite space parities and $CP$-parities.

Having identified the eigenstates by their approximate $CP$-parities
we can study time distributions of the cascade (1),(2) to determine
experimentally which of states is heavier or lighter (detailed
expressions see in ref.\cite{az}). Of course, these two
identifications, by $CP$-parity and by mass, should be equivalent.
The only problem is how to relate them. Cascade decays just open
such a possibility. Any other way has not been suggested yet.

3. Here appears one more interesting point. Let us compare
violations of space parity and $CP$-parity. In decays caused by
strong interactions, where the space parity is violated only to a
very small extent, we can unambiguously determine parity of an
unstable state (e.g., resonance) as the parity of any of its final
states. In weak decays the situation is different: various final
states may have various parity values. As a result, for example,
the space parity of the charged kaon is opposite when determined
from two- or three-pion final states. This is possible due to the
intrinsically large parity violation in weak interactions.
Nevertheless, manifestations of the parity violation in a concrete
weak decay may be very small or even not observable at all due to
some special reasons (e.g., absence of pseudoscalars because of
kinematical reasons).

$CP$-violation is usually believed to be intrinsically small, or
even "superweak"~\cite{W}. However, till now it has been definitely
observed only in decays of neutral kaons and only in three of their
decay modes ($\pi^0\pi^0,\, \pi^+\pi^-,\,\pi^+\pi^- \gamma$). All
of them give evidence for small $CP$-violation (in particular, they
all ascribe the same $CP$-parities to $K_S,\,K_L$). However, it might
be due to some "kinematical" reasons since all the kaon decays
correspond mainly to the same weak quark transition $s\to uW^-$
(other transitions, e.g., penguins $s\to dg$, seem to be relatively
small).

$B$-mesons have much richer variety of decay modes, which can be
induced by various underlying quark transitions. Therefore, they
allow much deeper studies of $CP$-violation. In particular, there
are different cascades, similar to (1), but induced by different
quark processes, e.g.,
\begin{equation} B_d(\overline B_d)\to \rho^0 K^0(\overline K^0)\,.
\end{equation}
They might give other $CP$-parities for $B_H,\,B_L$, than cascade
(1). In this sense the approximate $CP$-parity of, say, $B_L$ might
appear decay-mode dependent. In direct similarity to the case of
$P$-violation, this would mean that $CP$-violation is really
intrinsically large and appears to be small in kaon decays only
"kinematically". We see~\cite{az}, that cascade decays of $B $-mesons
suggest an experimental way appropriate to discover such a possibility
\footnote{I thank I.Dunietz for discussions which imply that present
data on the CKM-matrix may lead just to the opposite $CP$-parities of,
say, $B_L$ as determined from cascades (1) and (6).}.

4. Now we consider direct non-leptonic decays to final states of
definite $CP$-parity. The simplest example for kaons is the decay
\begin{equation}
K^0(\overline K^0)\to\pi^+\pi^-\,. \end{equation}
Its well-known time distribution contains the oscillating term
proportional to $\cos(\phi_{+-}- \Delta m_K t)$, where $\phi_{+-}$
is the phase of the $CP$-violating parameter $\eta_{+-}$. Note that
the famous parameters $\eta$'s are inevitably defined in terms of
eigenstates with definite $CP$-parities (at least approximate), so
to have $|\eta| < 1$. The states being identified as heavy or light
(or in any other way) may be used only if their $CP$-parities are
known. So, $\Delta m_K$ in the time distribution is the mass
difference of kaon eigenstates of definite (m.b., approximate) $CP
$-parity and, {\it a priori}, may have an arbitrary sign. We see,
therefore, that the sign of $\phi$ can be measured only in respect
to the sign of $\Delta m_K$. In other words, unambiguous measurements
of $\eta$'s are possible only when we know what are $CP$-parities of
heavier and lighter eigenstates. As we mentioned above, for kaons it
was really done in special experiments.

The situation is the same for neutral $B$ (or $D$) mesons~\cite{ars}.
The only difference is that the sign of, say, $\Delta m_B$ cannot be
measured in the same way as for kaons. It can be determined only in
cascade decays with double-flavor oscillations. As a result,
experiments with cascade decays are a necessary step for complete
unambiguous measurements of $CP$-violating parameters in decays of
neutral $B$ mesons~\cite{ars} (see also ref.\cite{az2}; similar
discussion in terms of the CKM-angle $\beta$ is given in
ref.\cite{bk}).

Furthermore, the double-time distribution in cascade decays (1),(2)
(over $t_B$ and $t_K$) contains the $CP$-violating parameter of a
new kind.  It can be presented as
\begin{equation}
\eta^{J/\psi}_{BK}=\frac{1+\lambda^{J/\psi}_{BK}}{1-
\lambda^{J/\psi}_{BK}}\,, \end{equation}
where~\cite{az}
\begin{equation}
\lambda^{J/\psi}_{BK}=\frac{1-\epsilon_B}{1+\epsilon_B}\;
\frac{1+\epsilon_K}{1-\epsilon_K}\;\frac{<J/\psi
\overline K^0|\overline B^0>}{<J/\psi K^0|B^0>}\,.
\end{equation}
This parameter determines the relative value of contributions of
"suppressed" branches (5) to the total decay amplitude in comparison
with "unsuppressed" branches (4). Note that the exact $CP
$-conservation would lead to $\lambda^{J/\psi}_{BK}=-1$ and to the
vanishing of $\eta^{J/\psi}_{BK}$. Similar parameters should appear
in other cascade decays as well. An interesting specific property of
these new parameters is that they reveal interplay of $CP$-violation
in both neutral $B$- and neutral $K$-mesons (e.g., $\eta^{J/\psi
}_{BK}$ could not vanish even if $CP$ were conserved for $B$-mesons,
but violated in a usual way for kaons). Therefore, their measurements
may lead to deeper insight into the problem of $CP$-violation.

5. Let us discuss time dependence of the cascade decays. The familiar
opinion is that sequential decays (i.e. having several consequent
stages) have factorisable time dependence determined by
multiplication of time dependencies for each stage.
This is not true for the cascade (1),(2). Its amplitude~\cite{az}
can be presented in the form
\begin{equation} A_{B(0)\to final\,state}(t_B,\,t_K)\propto A_{+L}
+ A_{-S} +\eta^{J/\psi}_{BK} (A_{+S}+A_{-L})\,, \end{equation}
where the first two terms correspond to amplitudes of branches (4),
while the last two terms come from branches (5). Double-time
dependence of the amplitude for each branch is factorisable in $t_B$
and $t_K$. However, the dependencies are different for all the
branches, so the total amplitude is non-factorisable. The decay
yield is surely non-factorisable as well.

Nevertheless, the double-time dependence is really not very
complicated~\cite{az}. As the function of $t_K$, in similarity
with decays (2), it generally combines four terms:
$$ \exp(-\Gamma_S t_K)\,,\;\;\exp(-\Gamma_L t_K)\,,\;\;
\exp\left(-\frac{\Gamma_S+\Gamma_L}{2}t_K\right)\cos(\Delta m_K
t_K)\,,\;\; \exp\left(-\frac{\Gamma_S+\Gamma_L}{2}t_K\right)\sin(
\Delta m_K t_K)\,.$$
Their coefficients, in difference with decays (2), are not
constant.  They are functions of $t_B$, each one combining again
four terms:
$$ \exp(-\Gamma_+ t_B)\,,\;\;\exp(-\Gamma_- t_B)\,,\;\;
\exp\left(-\frac{\Gamma_++\Gamma_-}{2}t_B\right)\cos(\Delta m_B
t_B)\,,\;\; \exp\left(-\frac{\Gamma_++\Gamma_-}{2}t_B\right)\sin(
\Delta m_B t_B)\,.$$
Coefficients of these combinations (see ref.\cite{az}) depend on
the initial $B$-meson state and on properties of decays (1) and (2).
As the above discussion shows, the total amplitude (10) is sensitive
to the sign of $\Delta m_B$, if the sign of $\Delta m_K$ is fixed
and the eigenstates $B_{\pm}$ are identified so that $|\eta^{J/
\psi}_{BK}|<1$ (or, equivalently, Re$\,\lambda^{J/\psi}_{BK}<0$).

Non-factorisable structure of the amplitude (10) leads to one more
interesting consequence. Integration over one of lifetimes, $t_B$
or $t_K$, produces single-time distributions of the cascade, which
"remember" many properties of its double-time distribution.
Experimentally, it looks reasonable to integrate over $t_B$ (because
of very small mean lifetime of $B$-mesons) and study distribution
over the secondary-decay time $t_K$ to search for effects of
double-flavor oscillations~\cite{az,ars}. Principally, it might be
sufficient, e.g., for determination of approximate $CP$-parities of
$B_H, B_L$ (or, equivalently, the sign of $\Delta m_B$ at fixed
approximate $CP$-parities of eigenstates).

6. Let us discuss possibilities to accomplish experiments on
double-flavor oscillations. They look to be difficult experiments
indeed!

First of all, the present experimental data~\cite{pdt} for branching
ratios at various steps of the cascade (1),(2), appended by tagging
$J/\psi \to l^+l^-$, provide the small effective branching
\begin{equation}
({\rm Br})_{eff}\approx5\cdot 10^{-5}  \end{equation}
for any of secondary decays (2). And this is not the end of the story.

A necessary element for observations of double-flavor oscillations is
the interference of both $B_+, B_-$ (or $B_H, B_L$) and $K_S, K_L$.
Eigenstates of neutral $B$-mesons in the decay (1) have really the
same mean lifetimes and interfere at any $t_B$. Contrary to them,
$K_S$ and $K_L$ effectively interfere only in a limited interval
of $t_K$, thus producing, besides of small branching (11), some
additional smallness (its estimates see below).

As a result, the necessary experimental statistics seems to be quite
inaccessible at any of the projected \mbox{$B$-factories}
(corresponding distributions were, nevertheless, studied in
ref.\cite{ds}). LHC or some other hadron facility may be promising,
but statistics of the dedicated detector LHC-B being planned to
work at moderate luminosity and produce about $10^8$
$B$'s/year~\cite{lhc} looks insufficient as well~\cite{ars}. The
desired effects might be searched for either by other LHC detectors
or by LHC-B working at full luminosity.

An essential question for planning future experiments is which of
kaon decays (2) in the cascade (1),(2) would be preferable. All the
papers~\cite{ars,bk,ds} observe some evidence that the semileptonic
decays might be more favorable than two-pion ones. We can explain
this here as follows.  The smallness of $K_S,K_L$ interference in
each particular two-pion mode may be characterized by the small
parameter $|\eta_{\pi \pi}|\approx2.26\cdot10^{-3}$. Note that the
two-pion yield in decays of pure $K_L$, as compared to pure $K_S$,
is suppressed by $|\eta_ {\pi\pi}|^2$.

At first sight, $K_S, K_L$ interference in semileptonic decays
should not have any smallness since their decay amplitudes are the
same (or differ only in sign). Some effective smallness exists,
nevertheless.  Yield of semileptonic decays of pure $K_S$, in
comparison with pure $K_L$, would be suppressed by the factor
$\tau_S/\tau_L \approx1.7 \cdot10^{-3}$. Interference is linear in
each amplitude, so, the net (not at particular $t_K$!) effects of
$K_S, K_L$ interference in the semileptonic channels should be
characterized by the parameter $(\tau_S/\tau_L)^{1/2}\approx4.2
\cdot10^ {-2}$, which is also small, but about 20 times higher than
$|\eta_{\pi\pi}|$. This simple estimate quantitatively coincides
with the result of ref.~\cite{ars} obtained in a different way. It
influences requirements to necessary experimental statistics and
explains why semileptonic decays of the secondary kaon might be more
efficient than two-pion decays to look for double-flavor
oscillations. There is, however, a problem of experimental
efficiencies which should be studied separately for each particular
detector.

In summary, briefly explained here are the origin and some features
of coherent double-flavor oscillations, a new phenomenon which is
specific for decays of heavy neutral flavored mesons. Its
observation would provide unique important information on properties
of heavy mesons and better understanding of $CP$-violation in their
mixing and decays. Experiments searching for such oscillations are
hard, but look promising at hadron facilities.

Correspondence with B.Kayser during my work at this written version
of the talk was helpful to achieve better formulations of my ideas.


\begin{thebibliography}{99}
  \bibitem{aJ}
      Ya.\,I.\,Azimov, Pis'ma ZhETF 50 (1989) 413 [translation:
      JETP Lett. 50 (1989) 447].
\vspace{-2.5mm}
  \bibitem{az}
      Ya.\,I.\,Azimov, Phys.Rev. D42 (1990) 3705.
\vspace{-2.5mm}
  \bibitem{ad}
      Ya.\,I.\,Azimov, I.\,Dunietz, Phys.Lett. 395B (1997) 334;
      e-print hep-ph/9612265.
\vspace{-2.5mm}
  \bibitem{epr}
      A.\,Einstein, B.\,Podolsky, N.\,Rosen, Phys.Rev. 47 (1935) 777.
\vspace{-2.5mm}
  \bibitem{ks} B.\,Kayser, L.\,Stodolsky, preprint MPI-PhT/96-112;
      e-print hep-ph/9610522.
\vspace{-2.5mm}
  \bibitem{pdt-old} Particle Data
      Group, Phys.Lett. 170B (1986) 132.
\vspace{-2.5mm}
  \bibitem{pdt}
      Particle Data Group, Phys.Rev. D54 (1996) 1.
\vspace{-2.5mm}
  \bibitem{dan}
      Dan-di Wu, Phys.Rev. D40 (1989) 806.
\vspace{-2.5mm}
  \bibitem{ars}
      Ya.\,I.\,Azimov, V.\,L.\,Rappoport, V.\,V.\,Sarantsev,
      Z.Physik A356 (1997) 437;\\ e-print hep-ph/9608478.
\vspace{-2.5mm}
  \bibitem{W}
      L.\,Wolfenstein, Phys.Rev.Lett. 13 (1964) 562.
\vspace{-2.5mm}
  \bibitem{az2}
      Ya.\,Azimov, talk at the 2nd Intern.Conf. on
      $B$-physics and $CP$-violation, Honolulu, Hawaii,\\ March
      24-27, 1997 (to appear in the Proceedings); e-print
      hep-ph/9706463.
\vspace{-2.5mm}
  \bibitem{bk}
      B.\,Kayser, talk at the Moriond Workshop on Electroweak
      Interactions and Unified Theories,\\ Les Arcs, France, March
      1997 (to appear in the Proceedings); e-print hep-ph/9709382.
\vspace{-2.5mm}
  \bibitem{ds}
      G.\,V.\,Dass and K.\,V.\,L.\,Sarma, Intern.Journ.Mod.Phys. A7
      (1992) 6081; (E) A8 (1993) 1183.
\vspace{-2.5mm}
  \bibitem{lhc}
      LHC-B, {\em Letter of intent}. CERN/LHCC 95--5, August 1995.
\vspace{-2.5mm}

\end{thebibliography}
\end{document}